\documentclass[aps,showpacs,twocolumn]{revtex4}
\usepackage{amsfonts}
\usepackage{amsmath}
\usepackage{amssymb}
\usepackage{graphicx}
\usepackage{bm}

\input{tcilatex}

\begin{document}

\title{Collective thermo-electrophoresis of charged colloids}
\author{Arghya Majee and Alois W\"{u}rger}
\affiliation{Laboratoire Ondes et Mati\`{e}re d'Aquitaine, Universit\'{e} Bordeaux 1 \&
CNRS, 351 cours de la Lib\'{e}ration, 33405 Talence, France}

\begin{abstract}
Thermally driven colloidal transport is to a large extent due to the
thermoelectric or Seebeck effect of the charged solution. We show that,
contrary to the generally adopted single-particle picture, the transport
coefficient depends on the colloidal concentration. For solutions that are
dilute in the hydrodynamic sense, collective effects may significantly
affect the thermophoretic mobility. Our results provide an explanation for
recent experimental observations on polyelectrolytes and charged particles,
and suggest that for charged colloids collective behavior is the rule rather
than the exception.
\end{abstract}

\maketitle

\section{Introduction}

Transport in macromolecular or colloidal dispersions is mainly driven by
interface forces \cite{Der87,And89,Squ05,Boc10}. Because of the rather
short-ranged flow pattern induced in the surrounding fluid, these forces do
not result in hydrodynamic interactions, in contrast to diffusion and
sedimentation. As a consequence, nearby beads hardly see each other, and
their transport velocity is independent of concentration \cite{Ree75}; for
the same reason, free-solution electrophoresis of polyelectrolytes does not
depend on the molecular weight \cite{Gro92,Vio00}.\ Similar results have
been obtained for thermal diffusion of high polymers \cite%
{Gid76,Bro81,Wie04,Wue07}.

Recent experiments on thermophoresis in charged colloids, however, dress a
rather different picture and indicate that the single-particle description
fails in several instances: Contrary to expectation, the transport velocity
due to a temperature gradient,%
\begin{equation}
u=-D_{T}\nabla T,  \label{1}
\end{equation}%
was found to depend on the volume fraction of particle dispersions and on
the chain length $N$ of macromolecular solutions: (i) Data on sodium
polystyrene sulfonate (NaPSS) \cite{Iac06} and single-stranded DNA \cite%
{Duh06} at constant polymer content but variable $N$, reveal that the
mobility $D_{T}$ becomes smaller for larger molecules;\ e.g., in the range
from 50 to 48000 base pairs, that of DNA decreases by a factor of 5. These
findings are obtained at low concentration where the molecular mean distance
is much larger than the gyration radius. (ii) Regarding particle
suspensions, experiments on 70-nanometer silica beads \cite{Gho09} and 26-nm
latex spheres \cite{Put05} in a weak electrolyte, show that at a volume
fraction of 2\%, $D_{T}$ is significantly reduced with respect to the
zero-dilution value.

In the present work we show that these experimental findings arise from an
interaction mechanism that has been overlooked so far, i.e., the collective
thermoelectric response of the composite system. By treating the salt ions
and the dispersed colloid on an equal footing, we find that both the
thermoelectric field and the mobility $D_{T}$ vary with the colloidal
concentration. Depending on the electrolyte strength and the valency of the
macroions, collective effects may occur at low dilution, that is, for
particle dispersion with negligible pair potential and polymer solutions
where neighbor chains do not overlap.

Thermally driven motion of charged colloids is very sensitive to the solvent
composition. From previous work it emerges that two rather different
mechanisms contribute to the velocity \cite{Put05,Wue08},

\begin{equation}
u=-\mu _{T}\nabla T+\mu E.  \label{1b}
\end{equation}%
The first term arises from the local particle-solvent interactions in a
non-uniform temperature. As first pointed out by Ruckenstein \cite{Ruc81},
the temperature gradient deforms the electric double layer and induces a
pressure gradient opposite to $\nabla T$. The resulting thermoosmotic
surface flow toward higher $T$ drives the particle to the cold side; the
overall picture is similar to electroosmotic effects in an electric field 
\cite{Wue10}. The coefficient $\mu _{T}\propto \varepsilon \zeta ^{2}/\eta T$
depends on the $\zeta $-potential, and the solvent permittivity $\varepsilon 
$ and viscosity $\eta $; different prefactors occur in the limits of small
and large particles \cite%
{Duh06a,Dho07,Dho08,Mor08,Ruc81,Mor99,Bri03,Ras08,Wue10}. This form agrees
rather well with the observed salinity dependence \cite{Pia02}, yet fails in
view of the strong variation with $T$ reported for various systems \cite%
{Iac06,Duh06}, thus suggesting the existence of an additional, so far poorly
understood contribution to $\mu _{T}$.

The present work deals with the second term in (\ref{1b}), which accounts
for the Seebeck effect of the charged solution or, in other words, for
electrophoresis in the thermoelectric field $E$ with the mobility $\mu
=\varepsilon \zeta /\eta $. Due to their temperature dependent solvation
forces, ions migrate along or opposite to the thermal gradient.\ As a
consequence, surface charges develop at the cold and warm boundaries of the
vessel and give rise to a macroscopic electric field $E=-\psi \nabla T/T$;
see Fig. 1. The thermopotential parameter $\psi $ is related to the Seebeck
coefficient $S=-\psi /T$; for electrolytes $S$ attains values of several $%
100\mu $V/K, which is by one to two orders of magnitude larger than in
common metals \cite{Row06}.

\section{Thermophoretic mobility}

We consider a dispersion of negatively charged particles or macromolecules
of valency $-Z$ and concentration $n$, in a monovalent electrolyte solution
of ionic strength $n_{0}$ with a constant temperature gradient $\nabla T$.
According to the general formulation of non-linear thermodynamics the
currents of colloid and salt ions are linear functions of generalized forces 
\cite{Gro62}; the latter can be expressed through thermal and concentration
gradients. The current of colloidal macroions is given by

\begin{equation}
J=-D\nabla n+nu,  \label{4}
\end{equation}%
where the first term on the right-hand side accounts for normal diffusion
and the second one for transport with the drift velocity (\ref{1b}). 

The densities of small ions account for the counterions released by the
colloidal particles and the added salt. the salinity. The mobile ion currents

\begin{equation}
J_{\pm }=-D_{\pm }\left( \nabla n_{\pm }+2n_{\pm }\alpha _{\pm }\frac{\nabla
T}{T}\mp n_{\pm }\frac{eE}{k_{B}T}\right)  \label{3}
\end{equation}%
comprise normal diffusion with coefficients $D_{\pm }$, thermal diffusion
with the reduced Soret parameters $\alpha _{\pm }$, and electrophoresis with
the H\"{u}ckel mobility for monovalent ions. In (\ref{4}) and (\ref{3}) we
have added an electric field term; it is important to note that $E$ is not
an external field but arises from the kinetics of the mobile charges and is
proportional to the applied temperature gradient. A similar phenomenon
occurs in a non-uniform electrolyte, where the electric field is
proportional to the salinity gradient and to the difference of the ionic
diffusion coefficients $D_{\pm }$ \cite{And89,Pri87}.

The numbers $\alpha _{\pm }$ describe the drift of positive and negative
salt ions in a temperature gradient. The values for the most common ions
have been determined by Agar from thermopotential measurements of
electrolyte solutions \cite{Aga89}; our notation and Agar's
\textquotedblleft heat of transport\textquotedblright\ $Q_{\pm }^{\ast }$
are related through $\alpha _{\pm }=Q_{\pm }^{\ast }/2k_{B}T$. Typical
values range from $\alpha \approx 0$ for Li$^{+}$ to $\alpha \approx 3$ for
OH$^{-}$; those of the most common ions are given in Table I. 
\begin{table}[b]
\caption{Reduced Soret coefficient $\protect\alpha _{\pm }$ of several salt
ions at room temperature. The values of the heat of transport $Q_{\pm
}^{\ast }$ are taken from Ref. \protect\cite{Aga89}. The parameters $\protect%
\alpha _{\pm }$ are calculated from $\protect\alpha _{\pm }=Q_{\pm }^{\ast
}/2k_{B}T$. }%
\begin{tabular}{|l|c|c|c|c|c|c|}
\hline
Ion & H$^{+}$ & Li$^{+}$ & K$^{+}$ & Na$^{+}$ & OH$^{-}$ & Cl$^{-}$ \\ \hline
$Q_{i}^{\ast }$ (kJ/Mol) & $13.3$ & $0.53$ & $2.59$ & $3.46$ & $17.2$ & $%
0.53 $ \\ \hline
$\alpha _{i}$ & $2.7$ & $0.1$ & $0.5$ & $0.7$ & $3.4$ & $0.1$ \\ \hline
\end{tabular}%
\end{table}

\subsection{The steady state}

Eqs. (\ref{4}) and (\ref{3}) provide the currents as functions of the
generalized thermodynamic forces, that is, of the concentration and
temperature gradients \cite{Gro62}. We are interested in the steady state
characterized by 
\begin{equation}
J_{\pm }=0=J.  \label{6a}
\end{equation}%
For later use we give of a resulting relation for the electric field.
Inserting the drift velocity (\ref{1b}) and superposing the three equations (%
\ref{6a}) such that the concentration gradients result in the gradient of
the charge density, $\nabla \rho =e\nabla (n_{+}-n_{-}-Zn)$, and collecting
terms proportional to $E$ and $\nabla T$, one has 
\begin{equation}
E=e\frac{2n_{+}\alpha _{+}-2n_{-}\alpha _{-}-ZnT\mu _{T}/D}{\varepsilon
\kappa ^{2}}\frac{\nabla T}{T}+\frac{\nabla \rho }{\varepsilon \kappa ^{2}},
\label{6c}
\end{equation}%
with the shorthand notation $\kappa ^{2}=e^{2}(n_{+}+n_{-}+ZnT\mu
_{T}/D)/\varepsilon k_{B}T$.

In order to determine the four unknowns $\nabla n_{\pm }$, $\nabla n$, $E$,
the three equations (\ref{6a}) need to be completed by a fourth condition;
it is provided by Gauss law%
\begin{equation}
\mathrm{div}E=\rho /\varepsilon   \label{6b}
\end{equation}%
which relates $E$ and the charge density $\rho =e(n_{+}-n_{-}-Zn)$, and thus
closes the above set of equations.

\begin{figure}[b]
\includegraphics[width=\columnwidth]{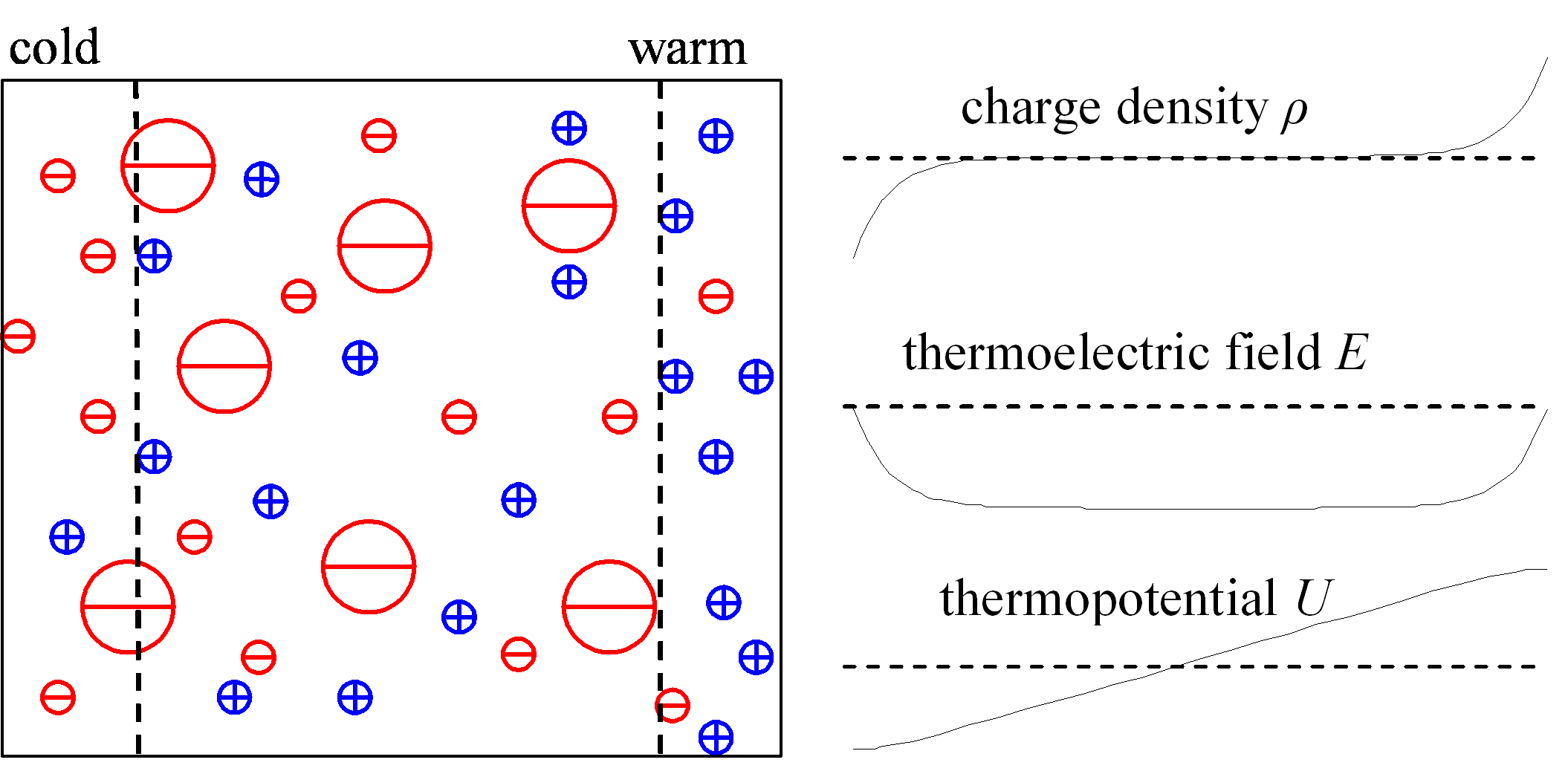}
\caption{Thermoelectric effect in a colloidal suspension of charged particles in
salt solution. In the example presented, the Soret parameters are such that
negative and positive ions accumulate at the cold and warm boundaries,
respectively. In the left panel, vertical dashed lines indicate the
thickness of the surface layers of about one Debye length $\protect\lambda $%
. This schematic view exaggerates the surface layers, which are much thinner
in real systems. The right panel shows the spatial variation of the net
charge density $\protect\rho $, the thermoelectric field $E$, and the
thermopotential $U$; dashed lines indicate the zero of the ordinate. Note
the non-zero surface charges at the cold and hot boundaries. The present
paper discusses the bulk behavior only, where $\protect\rho =0$ and where $E$
is constant. }%
\end{figure}

\subsection{Small-gradient approximation}

The above Eqs. (\ref{6a}) and (\ref{6b}) are non-linear in the
concentrations and thus cannot be solved as they stand. The salt and colloid
concentrations vary very little through the sample; the relative changes $%
\delta n/n$ and $\delta n_{\pm }/n_{\pm }$ between the hot and cold
boundaries are\ proportional to the reduced temperature difference $\delta
T/T$. Since in experiment, the ratio $\delta T/T$ is much smaller than
unity, we may safely replace the concentrations $n$ and $n_{\pm }$ in the
coefficients of (\ref{6c}) with constants $\bar{n}$ and $\bar{n}_{\pm }$;
the latter are defined as the colloidal and salt concentrations at $\nabla
T=0$.

Formally, this small-gradient approximations corresponds to neglecting terms
that are quadratic in the small quantities $\nabla n_{\pm }$, $\nabla n$, $E$%
, and $\nabla T$. This approximation has been used, more or less explicitly,
in previous works on the thermoelectric effect \cite{Gut49,Aga89} and in
recent applications in colloidal thermophoresis \cite{Put05,Wue08,Vig10}.
Moreover, various works on the osmotic flow driven by externally imposed
gradients of charged solutes resorts to the same approximation, albeit with
the salinity change $\nabla n_{0}$ instead of the temperature gradient \cite%
{And89,Pri87,Abe08}.

\subsection{Bulk thermoelectric field}

The above relations (\ref{6a})-(\ref{6b}) describe both the bulk properties
of a macroscopic sample and boundary effects such as the surface charges
that develop at the hot and cold boundaries; see Fig. 1. The thickness of
the surface layer is given by the Debye length and thus in the range between
one and hundred nanometers. This is much smaller than the sample size. Thus
we discard surface effects and discuss the bulk behavior only; a full
evaluation including surface effects is given in the Appendix.

In a macroscopic sample the net charge density vanishes because of the huge
electrostatic energy. With 
\begin{equation*}
\rho _{\text{bulk}}=0,
\end{equation*}%
Gauss' law imposes (\ref{6b}) imposes a constant electric field; its
explicit expression is readily obtained from (\ref{6c}) 
\begin{equation}
E=-\psi \frac{\nabla T}{T},  \label{9}
\end{equation}%
with the shorthand notation for the coefficient of $\nabla T/T$ 
\begin{equation*}
\psi =-e\frac{2\bar{n}_{+}\alpha _{+}-2\bar{n}_{-}\alpha _{-}-Z\bar{n}T\mu
_{T}/D}{\varepsilon \bar{\kappa}^{2}}
\end{equation*}%
and $\bar{\kappa}^{2}=e^{2}(\bar{n}_{+}+\bar{n}_{-}+Z\bar{n}T\mu
_{T}/D)/\varepsilon k_{B}T$.\ Note that we have used the small-gradient
approximation and replaced the colloidal and ion concentrations with their
mean values.

Although it is not always mentioned explicitly, the argument of zero bulk
charge density has been used in previous works on the Seebeck effect of
electrolytes \cite{Gut49,Aga89,Put05,Wue08,Vig10} and, more generally, for
colloidal transport in non-equilibrium situations involving thermal or
chemical gradients \cite{And89,Pri87,Abe08}.

\subsection{Zero-dilution limit}

We briefly discuss the case of a very dilute suspension where the colloidal
charges are negligible for the electrostatic properties. Putting $%
n\rightarrow 0$ in the electric field (\ref{9}) we have $\psi _{0}=-\left(
\alpha _{+}-\alpha _{-}\right) k_{B}T/e$ and 
\begin{equation*}
E_{0}=\left( \alpha _{+}-\alpha _{-}\right) \frac{k_{B}\nabla T}{e}.
\end{equation*}%
This expression has been used previously in \cite{Gut49,Put05,Wue08,Vig10}.
Note that the parameter $\kappa ^{-1}$ reduces to the usual exprerssion of
the Debye screening length.

Inserting the thermoelectric field $E$ in the drift velocity (\ref{1b}) and
comparing with (\ref{1b}) defines the thermophoretic mobility 
\begin{equation}
D_{T}^{0}=\mu _{T}+\frac{\varepsilon \zeta \psi _{0}}{\eta T}.  \label{2}
\end{equation}%
Not surprisingly it is independent of the colloidal concentration. The
parameter $\psi _{0}$ and the macroscopic thermopotential $U=\psi _{0}\delta
T/T$\ between the hot and cold vessel boundaries, are given by the steady
state of the electrolyte solution. With the numbers of Table I, one finds
the values $\psi _{0}=-15$ mV and $+70$ mV for NaCl and NaOH solutions,
respectively. Thus one expects $D_{T}^{0}$ to change its sign upon replacing
one salt by the other \cite{Wue08}. This is confirmed by a very recent study
on sodium dodecylsulfate (SDS) micelles, where the electrolyte composition
NaCl$_{1-x}$OH$_{x}$ was varied at constant ionic strength \cite{Vig10};
increasing the relative hydroxide content $x$ from 0 to 1 resulted in a
linear variation of the Soret coefficient $S_{T}$ and a change of sign at $%
x\approx \frac{1}{2}$ \cite{Vig10}.

\subsection{Collective effects on the electric field $E$}

Now we derive the main result of this paper, that is, the dependence of $E$
and $D_{T}$ on the colloidal concentration and, in the case of
polyelectrolytes, on its molecular weight. As two important parameters we
define the ratio of the colloidal charge density and the salinity,

\begin{equation}
\phi =\frac{Z\bar{n}}{n_{0}},  \label{2b}
\end{equation}%
and the ratio of colloidal electrophoretic mobility $\mu $ and diffusion
coefficient $D$, 
\begin{equation}
\xi =\frac{k_{B}T}{e}\frac{|\mu |}{D}.  \label{4a}
\end{equation}%
In the following we assume a negative surface potential. For typical
colloidal suspensions, the charge ratio is smaller than unity, $\phi \sim
0.1 $, whereas the parameter $\xi $ may exceed 10$^{2}$.

Rewriting the coefficient $\psi $ in (\ref{9}) in terms of the dimensionless
quantities $\phi $ and $\xi $, we have 
\begin{equation}
\psi =-\frac{2(1+\phi )\alpha _{+}-2\alpha _{-}-\phi T\mu _{T}/D}{2+\phi
+\phi \xi }\frac{k_{B}T}{e}.  \label{5}
\end{equation}%
Eq. (\ref{5}) shows how the thermoelectric field arises from the competition
of the Soret motion of the mobile ions and the colloidal solute. In the
low-dilution limit $\phi \rightarrow 0$ the first term in the numerator
reduces to $(\alpha _{+}-\alpha _{-})$ which corresponds to the response of
the electrolyte solution discussed in previous work \cite{Put05,Wue08,Vig10}.

The $\phi $-dependent term in the numerator becomes relevant where $\phi
\sim D/T\mu _{T}$ and, in particular, may change the sign of $\psi $ and
thus of the thermoelectric field. With typical values $T\mu _{T}\sim 10^{-9}$
m/s$^{2}$ one has $D/T\mu _{T}=10^{-3}$ for micron-size particles (and
polyelectrolytes of a gyration radius.of 1 $\mu $m), and $D/T\mu
_{T}=10^{-1} $ for 10-nanometer beads. This means that, at typical colloidal
densities, the thermoelectric field is essentially determined by the
macroions. The denominator in (\ref{5}) results in an overall decrease when
augmenting the colloidal concentration.

\subsection{Collective effects on the mobility\ $D_{T}$}

Now we determine the steady-state thermophoretic mobility. Plugging the
value of the electric field $E$ given in (\ref{9}) into the drift velocity (%
\ref{1b}) and comparing with (\ref{1}), we get 
\begin{equation}
D_{T}=\frac{D_{T}^{0}}{1+\frac{\phi }{2+\phi }\xi }.  \label{8}
\end{equation}%
where $D_{T}^{0}$ is defined by Eq. (\ref{2}) albeit with a modified
parameter \ 
\begin{equation}
\psi _{0}=-\frac{(1+\phi )\alpha _{+}-\alpha _{-}}{1+\phi /2}\frac{k_{B}T}{e}%
.  \label{7}
\end{equation}%
The mobility\ and its dependence on the ratio $\phi $ constitute the main
result of this paper. According to (\ref{2}), the sign of $D_{T}$ is
determined by the competition of the bare mobility $\mu _{T}$ and the
Seebeck term proportional to $\zeta \psi _{0}$. Since $\phi <1$ in most
cases, the numerator of (\ref{8}) is rather similar to the dilute case
discussed above (\ref{2}).

A much more striking variation arises from the denominator of (\ref{8}). For
typical values of the charge ratio $\phi \sim 0.1,$ collective effects set
in where $\frac{1}{2}\phi \xi \sim 1$, in other words where $\xi $ is of the
order of 20. For high polymers ($N=10^{3}...10^{6}$) and colloidal particles
in the range from ten nanometers to a micron, the parameter $\xi $ takes
values between 10 and $10^{3}$. This simple estimate suggests collective
effects to occur in many systems. A detailed comparison with experiment is
given in the following section.

In the limit of zero dilution $\phi \rightarrow 0$ one readily recovers the
expression (\ref{2}). The opposite case of a saltfree system leads to 
\begin{equation*}
D_{T}=\frac{D_{T}^{0}}{1+\xi },\ \ \ \ \ \ (\phi \rightarrow \infty )
\end{equation*}%
with $\psi _{0}$ determined by the counterions only. In view of the large
values of $\xi $ mentioned above, one expects a strong reduction of the
mobility in the salt free case.

\section{Comparison with experiment}

We discuss Eq. (\ref{8}) in view of recent experiments on colloidal
suspensions. At relevant values of the charge ratio ($\phi \sim 0.1$) the
numerator hardly differs from that of the dilute case. Thus in the following
we \ focus on the reduction of $D_{T}$ due to the denominator.

\subsection{Polyelectrolytes}

We start with experimental findings on polyelectrolytes at constant volume
fraction but variable molecular weight. In their study of 2 g/l of NaPSS in
a 100 mM/l NaCl solution, Iacopini et al. found a significant variation with
the chain length \cite{Iac06}: Fig. 2a shows the data measured at 30$%
{{}^\circ}%
$ C for molecules of 74, 160, and 360 repeat units, with an overall decrease
of the mobility by 40 percent. The same factor has been found in the
temperature range from 15 to 35$%
{{}^\circ}%
$ C.

The solid line represents collective effects arising from the denominator of
Eq. (\ref{8}). It has been calculated with the double-layer term in the
small-bead limit, assuming the monomer to be small as compared to the Debye
length ($R<\lambda $) \cite{Dho07,Dho08,Mor08}, 
\begin{equation*}
\mu _{T}=-\frac{d\varepsilon }{dT}\frac{\zeta ^{2}}{3\eta },
\end{equation*}%
and with the H\"{u}ckel-limit electrophoretic mobility $\mu =\frac{2}{3}%
\varepsilon \zeta /\eta $. Inserting the diffusion coefficient $%
D=k_{B}T/6\pi \eta R$ and the Bjerrum length $\ell _{B}=e^{2}/4\pi
\varepsilon k_{B}T$ in (\ref{4a}), we have 
\begin{equation}
\xi =\frac{e|\zeta |}{k_{B}T}\frac{R}{\ell _{B}}.  \label{14}
\end{equation}%
The theoretical curve of Fig.\ 2a is calculated with the parameters $\zeta
=-27$ mV, $nN=10$ mM/l, and $\phi =0.1$. Its variation arises only from the
gyration radius $R=\ell N_{K}^{1-\nu }N^{\nu }$; we have used the usual
exponent $\nu =\frac{3}{5}$, the size of a monomer $\ell =0.4$ nm, and the
number of monomers per segment $N_{K}=10$. The dashed line indicates the
mobility in the short-chain limit. The theoretical expression (\ref{8})
provides a good description of the reduction of $D_{T}$ with increasing
chain length.

\begin{figure}[b]
\includegraphics[width=\columnwidth]{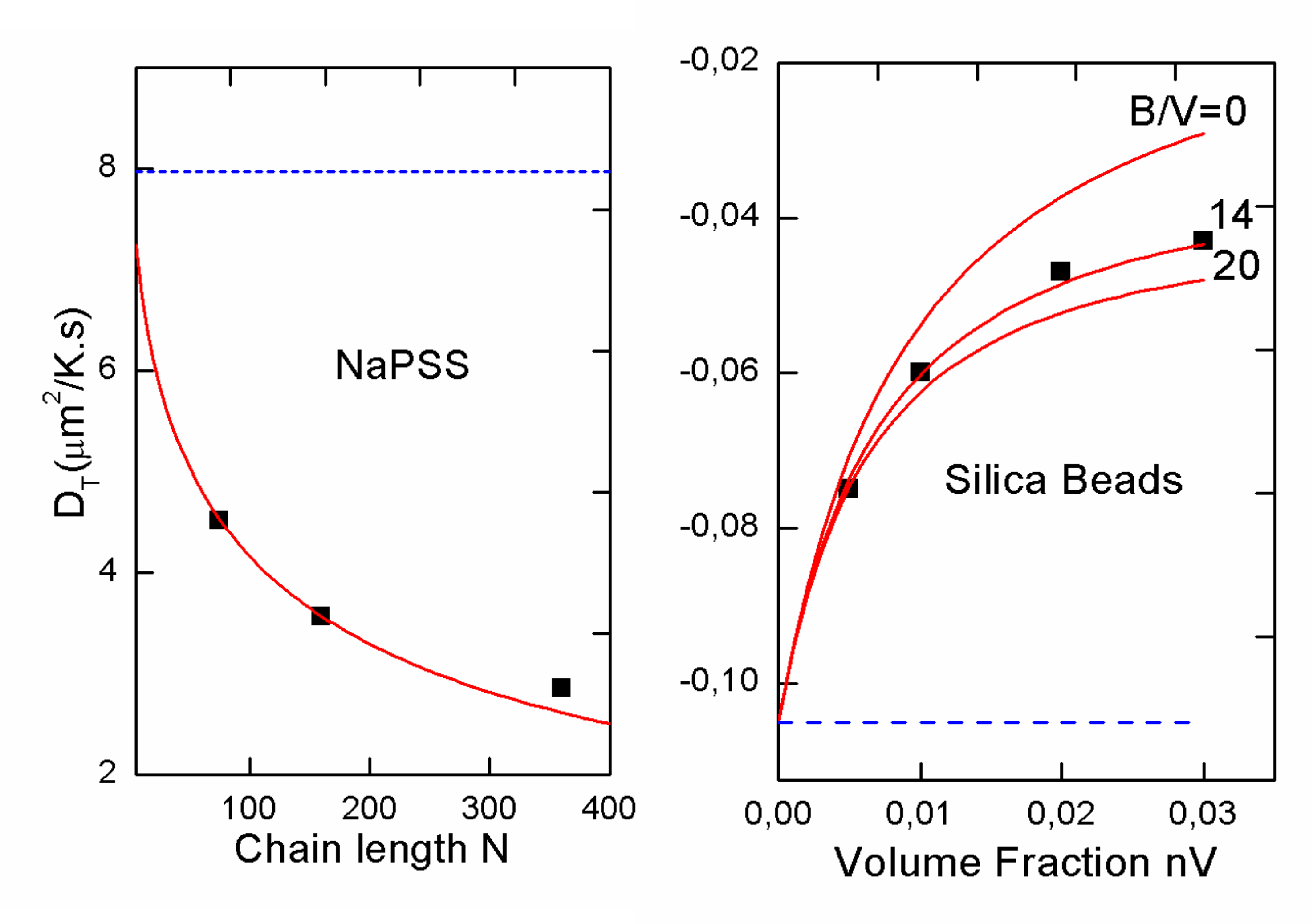}
\caption{ Comparison with
measured data. (a) Variation of $D_{T}$ with the chain length $N$ of a
polyelectrolyte at fixed volume fraction. The data on 2 g/l NaPSS in a 100
mM/l NaCl solution at 30$%
{{}^\circ}%
$ C are taken from Iacopini et al. \protect\cite{Iac06}. The solid line is
calculated from Eq. (\protect\ref{8}) with the parameters as given in the
main text. The dependence on $N$ arises from the gyration radius $R$. (b)
Volume fraction dependence of $D_{T}$ of a dispersion of 70-nm silica beads
in a solution of $30\protect\mu $M/l sulpho-rhodamine B.\ The data are from
Ghofraniha et al. \protect\cite{Gho09}; the fit curves are obtained from (%
\protect\ref{8}) and (\protect\ref{13}), with different values of the
reduced virial coefficient $B/V$, where $V=\frac{4}{3}\protect\pi R^{3}$ is
the particle volume.}%
\end{figure}

As a second example, DNA in 1 mM/l Tris buffer shows a similar behavior; its
mobility decreases by a factor of 5 over the range from $N=$50 to 48500 base
pairs per molecule \cite{Duh06}. The overall DNA content was kept constant, $%
Nn=50\mu $M/l, with a charge ratio $\phi =0.05$. Eqs. (\ref{8}) and (\ref{14}%
) provide a good fit to these data, albeit with a somewhat too small
exponent $\nu \approx 0.4$. In view of this discrepancy one should keep in
mind the rather complex electrostatic properties of polyelectrolytes.

The reduction observed for both NaPSS and DNA cannot be explained by
hydrodynamic effects. Interchain interactions are of little significance
because of the low dilution. Indeed, the effective volume fraction of the
polymer coils hardly attains a few percent, $nR^{3}\sim 10^{-2}$; thus
nearby chains do not overlap and leave both the viscosity and the diffusion
coefficient unchanged. Regarding hydrodynamic interactions of beads of the
same molecule, it is known that they enhance the electrophoretic mobility in
(\ref{1b}) and (\ref{2}) with increasing chain length.\ Yet this effect
occurs for short polyelectrolytes and saturates for chains longer than the
size of the screening cloud \cite{Mut96}; for the examples studied here, it
would enhance $D_{T}$ in the range $N<40$. We conclude that hydrodynamic
effects may ruled out as an explanation for the reduction shown in Fig. 2a.
Finally we discuss electrostatic single-particle effects. The
electrophoretic mobility in saltfree solution has been found to decrease
slightly at higher concentration, because of the increase of the overall
ionic strength and the shorter screening length \cite{Lob07,Med09}. In the
present case, however, the weight fraction of the polyelectrolyte is
constant, and so is the overall charge density. Thus the electrostatic
properties of the solution are the same for different chain lengths.

\subsection{Colloidal particles}

Now we discuss the concentration dependent mobility $D_{T}$ that has been
reported for dispersions of solid particles in weak electrolytes. Ghofraniha
et al. studied silica particles ($R=35$ nm) in a 30 $\mu $M/l solution of
the negatively charged dye sulpho-rhodamine\ B \cite{Gho09}. The data shown
in Fig. 2b reveal a significant decrease with the colloidal volume fraction;
at 3\% $D_{T}$ is reduced to less than half of the zero-dilution value. The
negative sign of the measured $D_{T}$ indicates that the thermoelectric
contribution $\varepsilon \psi _{0}\zeta /\eta T$ to (\ref{2}) overtakes the
Ruckenstein term \cite{Wue08} 
\begin{equation*}
\mu _{T}=\frac{\varepsilon \zeta ^{2}}{3\eta T}.
\end{equation*}%
The negative surface potential $\zeta $ implies that the thermopotential
parameter of the sulpho-rhodamine solution is positive, $\psi _{0}>0$.

The curves in Fig. 2b are calculated from (\ref{4a}) and (\ref{8}) with $%
\psi _{0}=10$ mV, which is comparable to common salts and weaker than the
values of NaOH and tetraethylammonium \cite{Put05,Vig10}. The rather small $%
D_{T}$ suggests that the particles are weakly charged; we use $Z=30$ and $%
\zeta =-10$ mV. The dashed line gives the mobility $D_{T}^{0}$ in the
zero-dilution limit, whereas the solid lines are given by (\ref{8}).

In addition to the explicit concentration dependence in terms of the
parameter $\phi $, one has to take into account that, even at moderate
colloidal volume fraction, the Einstein coefficient $D$ is not constant.\
Indeed, cooperative diffusion of charged particles arises from the
electrostatic pair potential $\Phi (\mathbf{r})$ and, to a lesser extent,
from hydrodynamic interactions \cite{Rus89}. To linear order in the
concentration, the virial expansion for the Einstein coefficient reads as%
\begin{equation}
D=D_{0}(1+2nB),  \label{13}
\end{equation}%
with the parameter 
\begin{equation*}
B=\frac{1}{2}\int dV\left( 1-e^{-\Phi /k_{B}T}\right) .
\end{equation*}%
For hard spheres the virial coefficient is given by the particle volume, $%
B=4V$ with $V=\frac{4}{3}\pi R^{3}$.\ The electrostatic pair potential
results in an effective interaction volume $V=\frac{4}{3}\pi (R+\frac{\chi }{%
2}\lambda )^{3}$, where $\lambda $ is the Debye length and $\chi $ a
numerical factor \cite{Pia02,Fay05,Vig10}; for small and highly charged
particles in a weak electrolyte, the repulsive forces may enhance the virial
coefficient by one or two orders of magnitude. On the other hand,
hydrodynamic interactions contribute a negative term $B/V\sim -6.5$ and
reduce the Einstein coeffient accordingly \cite{Rus89}. Our discussion of
the data of Ref. \cite{Gho09} is restricted to volume fractions up to 3\%;
at higher concentration the measured $D$ saturates and the linear
approximation ceases to be valid. In units of the particle volume $V$, the
measured virial coefficient reads $B/V=20$ \cite{Gho09}; the best fit of the
mobility data is obtained with $B/V=14$. This value is much larger than that
of hard spheres and thus indicates the importance of electrostatic
repulsion. The concentration of mobile charge carriers $n_{0}=30\mu $M/l
leads to a screening length of about $50$\ nm. With $\chi \sim 2$ in the
above expression for the effective volume, one finds a virial coefficient
close to the measured value. As an illustration of the effect of collective
diffusion on $D_{T}$, we plot Eq. (\ref{8}) for these three values: Though
the variation of $D_{T}$ with $B$ is not neglegible, it is significantly
weaker than that of the thermoelectric effect.

As a second experiment we mention data by Putnam and Cahill on latex beads
of radius $R=13$ nm in an electrolyte solution of 2mM/l ionic strength \cite%
{Put05}; varying the volume fractions from 0.7 to 2.2 wt\%, these authors
observed a reduction of $D_{T}$ by about 10 percent. With a valency of $%
Z\sim 50$ one finds that, at the highest particle concentration $n=4\mu $%
M/l, the charge ratio $\phi $ does not exceed 10 percent.

Finally we address the concentration dependence observed by Guarino and
Piazza for the Soret coefficient $S_{T}=D_{T}/D$ of SDS micelles \cite{Pia02}%
. Its decrease with the SDS content, is well described by collective
diffusion according to (\ref{13}). In a very recent measurement, Vigolo et
al. vary the electrolyte composition NaCl$_{1-x}$OH$_{x}$ and thus the
thermal diffusion parameter of the anion in (\ref{7}), $\alpha
_{-}=(1-x)\alpha _{\text{Cl}}+x\alpha _{\text{OH}}$ \cite{Vig10}. The
observed linear dependence of $S_{T}$ on $x$ confirms the crucial role of
the thermopotential. Unfortunately there are no mobility data for micelles;
thus at present it is not possible to determine whether their $D_{T}$ is
subject to collective effects similar to those of polyelectrolytes and solid
beads.

\section{Summary and conclusion}

In summary, charged colloids in a non-uniform temperature show collective
transport behavior mediated by the Seebeck effect of both colloidal and salt
ions. For large particles and macromolecules, cooperative effects set in at
rather low concentration, where hydrodynamic interactions are absent and
where the charge ratio $\phi $ is much smaller than unity. The criterion for
the onset of collective behavior, $\phi \xi \sim 1$ in (\ref{8}), involves
the ratio of the electrophoretic mobility and the Einstein coefficient; by
contrast, the criterion for cooperative diffusion, $Bn\sim 1$, depends on
the pair potential of the solute particles. The discussed examples suggest
that the collective thermoelectric effect is generic for colloids at
ordinary concentrations. This issue could be relevant for microfluidic
applications of thermophoresis.

We conclude with a remark on the thermoelectric field given in Eq. (\ref{5}%
). Both its magnitude and its sign can be tuned by chosing the appropriate
electrolyte and adjusting the charge ratio. With a thermal gradient of less
than one Kelvin per micron, $E$ may attain values of 100 V/m. Thus the
thermoelectric effect could be used for applying electric fields in
microfluidic devices. Local laser heating would permit to realize almost any
desired spatiotemporal electric-field pattern.\ 

Helpful and stimulating discussions with D.G.\ Cahill, R. Piazza, D.\ Braun,
and N.\ Ghofraniha are gratefully acknowledged.

\section{Appendix}

The thermoelectric field (\ref{9}) has been derived by using the charge
neutrality of the bulk of a macroscopic sample. Here we give a derivation
based on the \ steady state, Gauss' law, and the electrostatic boundary
conditions. Resorting to the small-gradient approximation, we replace the
coefficients in (\ref{6c}) by their mean values and thus have 
\begin{equation*}
E=-\psi \frac{\nabla T}{T}+\frac{\nabla \rho }{\varepsilon \ \bar{\kappa}^{2}%
}.
\end{equation*}%
From Gauss's law (\ref{6b}) one has $\nabla \rho /\varepsilon =\nabla ^{2}E$
and thus obtains a differential equations for the thermoelectric field $E$
with a constant inhomogeneity $-(\psi /T)\nabla T$, 
\begin{equation*}
E-\frac{\nabla ^{2}E}{\bar{\kappa}^{2}}=-\psi \frac{\nabla T}{T}.
\end{equation*}
The solution $E=E_{\text{inh}}+E_{\text{h}}$ consists of two contributions.
The inhomogeneous term $E_{\text{inh}}=-(\psi /T)\nabla T$ accounts for the
macroscopic Seebeck effect. The remaining one $E_{\text{h}}$ is related to
surface charges at the cold and hot boundaries of the sample. The
homogeneous equation $\nabla ^{2}E_{\text{h}}=\bar{\kappa}^{2}E_{\text{h}}$
is solved by the exponential function,%
\begin{equation*}
E_{\text{h}}=A_{+}e^{\bar{\kappa}z}+A_{-}e^{-\bar{\kappa}z},
\end{equation*}%
where $z$ is the coordinate in the direction of the temperature gradient.
Its range is $-\frac{1}{2}L\leq z\leq \frac{1}{2}L$ with the sample size $L$.

The electrostatic boundary conditions require that the electric field
vanishes at $z=\pm \frac{1}{2}L$. Putting $E=0$ and solving for the
coefficients of $E_{\text{h}}$, one readily finds $A_{\pm }=-\frac{1}{2}E_{%
\text{inh}}/\cosh (\bar{\kappa}L/2)$ and the thermoelectric field 
\begin{equation*}
E=-\frac{\psi }{T}\nabla T\left( 1-\frac{\cosh (\bar{\kappa}z)}{\cosh (\bar{%
\kappa}L/2)}\right) .
\end{equation*}%
Both $E$ and the corresponding charge density $\rho $ are illustrated in the
right panel of Fig. 1. The field vanishes at the boundaries and reaches its
constant bulk value (\ref{9}) within a few screening lengths $\bar{\kappa}%
^{-1}$. The parameter $\bar{\kappa}^{-1}$ takes values in the range between
1 and 100 nanometers and thus is much smaller than the size of sample $L$.
Even in microfluidic devices, $\bar{\kappa}L$ is in general larger than 10$%
^{3}$.

In real systems a more complex picture may emerge from the surface roughness
of the boundaries, the solute size, and surface charges of other origin.
Note that such additional effects do not affect the bulk electric field (\ref%
{9}) and thus are irrelevant for the results of this paper.


\begin{thebibliography}{99}
\bibitem{Der87} B.V. Derjaguin, N.V. Churaev, V.M. Muller, \textit{Surface
Forces}, Plenum Press, New York (1987)

\bibitem{And89} J.L.\ Anderson, Ann. Rev. Fluid Mech. \textbf{21}, 61
(1989)\ 

\bibitem{Squ05} T. M. Squires, S. R. Quake, Rev. Mod. Phys. \textbf{77}, 977
(2005)

\bibitem{Boc10} L.\ Bocquet, E.\ Charlaix, Chem.\ Soc. Rev. \textbf{39},
1073 (2010)

\bibitem{Ree75} L.D.\ Reed, F.A.\ Morrison Jr., J.\ Coll. Interf. Sci. 
\textbf{54}, 117 (1976)

\bibitem{Gro92} P.D. Grossman, J.C. Colburn (Ed.), \textit{Capillary
Electrophoresis}, Academic Press, (1992)

\bibitem{Vio00} J.L.\ Viovy, Rev.\ Mod.\ Phys. \textbf{72}, 813 (2000)

\bibitem{Gid76} J.C. Giddings et al., Macromolecules \textbf{9}, 106 (1976)

\bibitem{Bro81} F.\ Brochard, P.-G.\ de Gennes, C.\ R. Acad.\ Sc. Paris, S%
\'{e}rie II \textbf{293}, 1025 (1981)

\bibitem{Wie04} S.\ Wiegand, J.\ Phys. Cond. Matt. \textbf{16}, 357 (2004)

\bibitem{Wue07} A.$\ $W\"{u}rger, Phys. Rev. Lett. \textbf{98}, 138301 (2007)

\bibitem{Iac06} S.\ Iacopini et al., EPJ E \textbf{19}, 59 (2006)

\bibitem{Duh06} S.\ Duhr, D.\ Braun, PNAS \textbf{103}, 19678 (2006)

\bibitem{Gho09} N. Ghofraniha, G. Ruocco, C. Conti, Langmuir \textbf{25},
12495 (2009)

\bibitem{Put05} S.A. Putnam, D.G.\ Cahill., Langmuir \textbf{21}, 5317 (2005)

\bibitem{Wue08} A. W\"{u}rger, Phys. Rev. Lett. \textbf{101}, 108302 (2008).

\bibitem{Ruc81} E.$\ $Ruckenstein, J. Colloid Interface Sci. \textbf{83}, 77
(1981)

\bibitem{Wue10} A. W\"{u}rger, Rep. Prog. Phys. \textbf{73}, 126601 (2010)

\bibitem{Mor99} K.I.\ Morozov, JETP \textbf{88}, 944 (1999)

\bibitem{Bri03} E.\ Bringuier, A.\ Bourdon, Phys. Rev. E \textbf{67}, 011404
(2003)

\bibitem{Duh06a} S.\ Duhr, D.\ Braun, Phys. Rev. Lett. \textbf{96}, 168301
(2006)

\bibitem{Dho07} J.K.G.\ Dhont et al., Langmuir \textbf{23}, 1674 (2007)

\bibitem{Dho08} J.K.G.\ Dhont, W.J. Briels, Eur. Phys. J. E \textbf{25}, 61
(2008).

\bibitem{Mor08} J. Morthomas, A. W\"{u}rger, Eur. Phys. J. E \textbf{27},
425 (2008)

\bibitem{Ras08} S.N.\ Rasuli, R. Golestanian, Phys. Rev. Lett. \textbf{101},
138301 (2008)

\bibitem{Pia02} R.~Piazza, A.~Guarino, Phys. Rev. Lett. \textbf{88}, 208302
(2002)

\bibitem{Row06} D.M. Rowe, \textit{Thermoelectrics Handbook---Macro to Nano}%
, CRC-Taylor \& Francis, Boca Raton, (2006).

\bibitem{Gro62} S.R. de Groot, P.\ Mazur, \textit{Non-equlibrium
Thermodynamics}, North Holland Publishing, Amsterdam (1962)

\bibitem{Pri87} D.\ C.\ Prieve, J.\ Chem. Soc. Faraday Trans. 2, \textbf{83}%
, 1287 (1987)

\bibitem{Gut49} G. Guthrie et al., J.\ Chem. Phys.\ \textbf{17}, 310 (1949)

\bibitem{Aga89} J.N. Agar et al., J.\ Phys.\ Chem. \textbf{93}, 2082 (1989)

\bibitem{Vig10} D. Vigolo, S.\ Buzzaccaro, R.\ Piazza, Langmuir (2010)

\bibitem{Abe08} B.\ Ab\'{e}cassis et al., Nature Mat. \textbf{7}, 785 (2008)

\bibitem{Fay05} S.\ Fayolle et al., Phys. Rev. Lett.\ \textbf{95}, 208301
(2005)

\bibitem{Mut96} M.\ Muthukumar, Electrophoresis \textbf{17}, 1167 (1996)

\bibitem{Lob07} V.\ Lobaskin, B.\ D\"{u}nweg, M.\ Medebach, T. Palberg, C.\
Holm, Phys.\ Rev. Lett. \textbf{98}, 176105 (2007)

\bibitem{Med09} M.\ Medrano, A.T.\ P\'{e}rez, L.\ Lobry, F.\ Peters,
Langmuir \textbf{25}, 12043 (2009)

\bibitem{Rus89} W. Russel,\ D. Saville, W. Schowalter, \textit{Colloidal
Dispersions}, Cambridge University Press (1989)
\end{thebibliography}
\end{document}